# Chiral standing waves and its trapping force on chiral particles

Tianhang Zhang[1,2,^,*], M.R.C. Mahdy[3,4,^,*], Shadman Sajid Dewan[3], Md. Nayem Hossain[3], Hamim Mahmud Rivy[3], Nabila Masud[3], Ziaur Rahman Jony[3]

[1] *Department of Electrical and Computer Engineering, National University of Singapore, 4 Engineering Drive 3, Singapore 117583*

[2] *CGG Service Singapore, 9 Serangoon North Ave 5, Singapore 554531*

[3] *Department of Electrical & Computer Engineering, North South University, Bashundhara, Dhaka 1229, Bangladesh*

[4] *Pi Labs Bangladesh LTD, ARA Bhaban, 39, Kazi Nazrul Islam Avenue, Kawran Bazar, Dhaka 1215, Bangladesh*

^ Equal contributions

* Corresponding authors: a0113469@u.nus.edu , mahdy.chowdhury@northsouth.edu

Up to now, in the literature of optical manipulation, optical force due to chirality usually coexists with the non-chiral force and the chiral force usually takes a very small portion of the total force. In this work, we investigate a case where the optical force exerted on an object is purely due to the chirality while there is zero force on non-chiral object. We find that a trapping force arises on chiral particles when it is placed in a field consisted of two orthogonally polarized counter-propagating plane waves. We have revealed the underlying physics of this force by modeling the particle as a chiral diploe and analytically study the optical force. We find besides chirality; the trapping force is also closely related to the dual 'electric-magnetic' symmetry of field and dual asymmetry of material. We also demonstrate that the proposed idea is not restricted to dipolar chiral objects only. Chiral Mie objects can also be trapped based on the technique proposed in this article. Notably, such chiral trapping forces have been found robust by varying several parameters throughout the investigation. This trapping force may find applications in identifying objects' chirality and the selective trapping of chiral objects.

# 1. Introduction

A focused laser beam can attract an object towards the area with highest field intensity. Based on this gradient force, optical tweezers was invented by Ashkin and his co-workers[1] in 1986. Since then it has been developed into a very powerful tool to trap varieties of materials in a microscopic system and revolutionized the field of micromanipulation[2]. Besides single beam optical tweezers, multiple traps were generated by utilizing interfering waves[3], time sharing techniques[4] or computer programmed spatial light modulators[5] to manipulate multiple objects at the same time. By integrating with microfluidic systems, utilizing near-field structures or taking advantages of non-diffraction beams, novel optical manipulations such as selective trapping[6], template trapping[7], sorting[8] and optical tractor beams[9, 10] were also achieved. Conventionally, selective trapping or sorting are done based on some physical properties of objects' i.e. sizes or refractive index. In recent years, another physical property of an object which is more related to its inner structures *i.e.* chirality, is receiving more and more attention in the area of optical manipulation due to its potential applications in biological research and pharmacy industry.

An object or a system is said to be chiral when its mirror image cannot be superimposed with itself; no matter how to orient them. This lack of mirror symmetry is quite common in chemistry, biology and pharmacy: most of amino acids, nucleosides are chiral compounds; proteins, DNAs/RNAs are chiral structures and more than half of drugs are chiral products [11, 12]. Chiral light-matter interactions have found applications in optical sensing of chiral molecules [13-15] and designing chiral metamaterials [16, 17]. Investigation of optical force on chiral objects may bring up a new method to manipulate and separate chiral molecules non-invasively as a complementary of current used chemical methods. In the light-matter interaction, when either the light whose chirality associated with its polarization or the objects whose chirality associated with its structure are chiral, chirality would bring up new phenomenon. In such chiral interactions, discriminations were found between the left handed and the right-handed objects. Optical force on chiral spheres or chiral molecules were theoretically studied and methods were proposed to separate chiral objects with opposite handedness in [18-20]. Chiral light-matter interactions were also linked to optical binding[21] and recently developed optical pulling force[22]. Moreover, a force whose direction is perpendicular to the light propagating direction is found on a chiral object when it is placed above a slab[23] or in an evanescent wave[24] due to the asymmetrical energy flow in the lateral direction. Chiral discriminations were also experimentally demonstrated in both gradient force based selective trapping[25] and radiation pressure based optical sorting[26]. However, in the studies above, the optical force due to the chirality usually coexists with the non-chiral force and the chiral force usually takes a very small portion of the total force. In this paper, we investigate a case when the optical force exerted on an object is purely due to the chirality while there is zero force on non-chiral object. We have found that a trapping force arises on chiral particles when it is placed in a field consisted of two orthogonally polarized counter-propagating plane waves. We have revealed the underlying physics of this force by modeling the particle as a chiral diploe and analytically study the optical force. We have found besides chirality; the trapping force is also closely related to the dual 'electric-magnetic' symmetry of field and dual asymmetry of material. By increasing the energy of incident light (i.e. source frequency), number of the trapping positions can be increased. We also demonstrate that the proposed idea is not restricted to dipolar chiral objects only. Chiral Mie objects can also be trapped based on the technique proposed in this article. Notably, such chiral trapping forces have been found robust by varying several parameters throughout the investigation. The number of trapping positions remain the same (as the air background) when the background medium is water. As most of the experimental configurations usually use the water medium as background, we may conclude that: the proposals presented in this article may remain valid in realistic experimental set-ups with water or other similar background environments. This trapping force may find applications in identifying objects' chirality and selective trapping chiral objects.

The paper is arranged as following. In section 2, we use the full-wave simulation to calculate the optical force on a spherical particle made of chiral materials when it is placed at different positions in the optical field. In section 3 and 4, we analytically writing the optical force on a chiral dipole in the field to reveal the origin of the force. In section 5, we extend further to more complex two-dimensional case where it is interesting to find even for the same optical field, objects with different chirality will 'see' different trapping potential landscapes. In section 6, we have shown the change in optical force due to change in the radius of the nanoparticle and the refractive index of the surrounding medium. Next section, we have investigated change in trapping position with respect to wavelength. Finally, conclusions are drawn.

## 2. Selective trapping force on chiral particles in chiral interference wave

The chiral object we use in the paper is modeled as a spherical particle made of chiral material defined by the constitutive relations[27].

$$D = \varepsilon_r \varepsilon_0 E + i\kappa/cH$$
$$B = -i\kappa/cE + \mu_r \mu_0 H \qquad (1)$$

$\varepsilon_0, \mu_0, \varepsilon_r$ and $\mu_r$ are permittivity, permeability of the free space and the relative permittivity and permeability of the material. c is the speed of light in vacuum. The chirality parameter $\kappa$ is used to describe the chirality of the structure. It takes positive, negative or zero value for right handed, left-handed and non-chiral particles.

The optical field is depicted in the inset of Fig.1a which consisted of two counter-propagating waves with orthogonal polarizations. Spherical particles with chirality parameter $\kappa = 0.5$, -0.5 and 0 are placed at different positions along light propagation direction when full wave simulation is done. For most of the cases, the incident wavelength is 600nm, the radius of the particle is 100nm and the permittivity of the particle $\varepsilon_r$ is set to be 5. But we have varied the radius, wavelength and the permittivity of the background medium in sections 6 and 7. The incident power is set to be 1mW/$um^2$ throughout the paper. The optical force on the particles is calculated by Maxwell Stress Tensor method and plotted against their positions in Fig.1b.

For the optical field depicted in Fig.1a, the intensity is uniform anywhere in space since the electric field of two orthogonally polarized light do not inference with each other. Thus, for a normal dielectric particle with $\kappa = 0$, the optical force is zero everywhere since there is no gradient force involved and radiation pressures generated by two beams exactly cancel each other. However, for the chiral particles with $\kappa = 0.5$ and -0.5, a force arises and presents an interference pattern along light propagation direction as if the particle is placed in a standing wave. The optical force exert on particles with opposite handedness are always in the same magnitude and opposite direction at certain position. By considering equilibrium positions, particles with opposite handedness are trapped at different locations separated by a quarter wavelength which is shown in Fig.1a. To address the underlying physics of this trapping force, we modeled the sphere as a small chiral dipolar particle and using scattering theory to analytically write the optical force on the particle in term of incident field in the following two sections.

## 3. Chirality of interfering waves

The chirality of an object can be easily observed from the geometry. In order to characterize the local chirality of a monochromatic field, the quantity of chirality density

$K = K_e + K_m = \frac{\varepsilon_0}{2} E \cdot \nabla \times E + \frac{\mu_0}{2} H \cdot \nabla \times H$ is introduced in [14, 15, 28]. *E* and *H* are the time-dependent electric and magnetic field of a monochromatic light. Making an analogy to the energy density and energy flow density in Poynting theorem, the chirality flow density is defined as $\Phi = \Phi_e + \Phi_m = \frac{1}{2} E \times (\nabla \times H) - \frac{1}{2} H \times (\nabla \times E)$ which satisfies the continuity equations $\frac{\partial K}{\partial t} + \nabla \cdot \Phi = 0$. K and Φ are both time independent, suppose in free space, by using Maxwell's equations we get their time average expressions.

$$\bar{K} = \frac{\omega}{2} \varepsilon_0 \mu_0 \text{Im}[E_0 \cdot H_0^*]$$

$$\bar{\Phi} = -\frac{\omega}{4} \text{Im}[\varepsilon_0 E_0 \times E_0^* + \mu_0 H_0 \times H_0^*] \qquad (2)$$

$E_0$ and $H_0$ are the standard complex expressions of the field. It is shown in [28] that the chirality density K and chirality flow density Φ are actually the energy and momentum operators of the field multiplied by the helicity h. The chirality flow density is also direct related to the SAM density s.[29]

The polarization describes how electric field moves in a single plane while chirality describes a more complicated physical picture that the field rotate around an axis and propagate along it at the same time. Herein we consider the optical field consisted of counter-propagating plane waves with orthogonal polarizations. The field was used as an example in [14] and [30] to illustrate the difference between polarization and chirality. We write the total field as

$$\tilde{E} = E_{inc} \exp(ikx) \hat{y} + E_{inc} \exp(-ikx) \hat{z} \qquad (3)$$

The ellipticities of the electric and magnetic field are $\sigma_e = -\sin(2kx)$ and $\sigma_m = \sin(2kx)$ respectively. These values oscillate between -1 and 1 indicating the polarizations oscillate between linear and circular along x direction. $\sigma_e$ and $\sigma_m$ are always opposite number indicating the electric field and the magnetic field rotate in opposite sense at any position.

Substituting the field into (1), we find the chirality density $\bar{K}$ is zero everywhere. The charity flow is

$$\bar{\Phi} = \bar{\Phi}_e + \bar{\Phi}_m = -\frac{1}{2} \omega \varepsilon_0 E_{inc}^2 \sin(2kx) + \frac{1}{2} \omega \mu_0 H_{inc}^2 \sin(2kx) \qquad (4)$$

Although the electric ($\bar{\Phi}_e$) and magnetic components ($\bar{\Phi}_m$) of the charity flow are non-zero, the total charity flow ($\bar{\Phi}$) is zero since both components contribute equally to the total flow and always flow in opposite directions. This equal contribution fulfills the principle of the dual 'electric-magnetic' symmetry of Maxwell equations in free space[31]. Being different from the field, matter is strongly dual-asymmetric. As a result, in the light-matter interaction, the electric and magnetic parts of the field may contribute very differently in term of momentum transfer. In the following section, we will show that it is this unequal contribution brings about the trapping force on chiral particles.

## 4. Mechanical action of field on chiral dipoles

In this section, we model the spherical particle made of chiral material in section 2 as chiral dipoles. To describe the relationship between the induce momentums and external field, chiral polarizability $\chi$ is introduced besides the complex electric polarizability $\alpha_e$ and magnetic polarizability $\alpha_m$. $\chi$ is a complex function of chirality parameter $\kappa$ and $\chi = 0$ for particles without charity. The induced electric ($p$) and magnetic moments ($m$) for a chiral dipole are as following:

$$p = \alpha_e E + \chi H$$
$$m = -\chi E + \alpha_m H \tag{5}$$

We use Equation (2) in [32] to calculate the time average optical force on a chiral dipole in free space.

$$\langle F \rangle = \frac{1}{2}\text{Re}[p(\nabla \otimes E^*) + m(\nabla \otimes H^*) - \frac{ck^4}{6\pi}(p \times m^*)] \tag{6}$$

E and H are the incident electric and magnetic field vector and k is the wave number. The symbol $\otimes$ represent the dyadic product and $A(\nabla \otimes B)$ can be written as $A(\nabla \otimes B) = (A \bullet \nabla)B + A \times (\nabla \times B)$. Equation (6) calculates the optical force on a small spherical particle within the range of validity of the dipolar approximation. When the size of particle becomes lager, the equation is no longer accurate since higher order modes come into play. The first, second and third term in the equation describe the contributions from electric dipole, magnetic dipole and the direct interaction between electric and magnetic dipoles respectively. We separate the force into the dipole part $F_{EM}$ (first two terms) and interaction part

F$_{int}$ (the third term). Substitute (5) into (6) and reorganize the terms, the force can be written as (7) and (8).

$$\langle F_{EM} \rangle = \frac{1}{4}(\text{Re}[\alpha_e]\nabla|E|^2 + \text{Re}[\alpha_m]\nabla|H|^2) + 2\omega(\mu_0 \text{Im}[\alpha_e] + \varepsilon_0 \text{Im}[\alpha_m])\bar{p} + \frac{c^2}{\omega}\text{Re}[\chi]\nabla\bar{K} - 2\text{Im}[\chi]\bar{\Phi}$$
$$+ \frac{2}{\omega\varepsilon_0}\text{Im}[\alpha_e]\nabla \times \bar{\Phi}_E + \frac{2}{\omega\mu_0}\text{Im}[\alpha_m]\nabla \times \bar{\Phi}_M - \text{Im}[\chi]\nabla \times \bar{p}$$
(7)

$$\langle F_{INT} \rangle = -\frac{ck^4}{6\pi}(2(\text{Re}[\alpha_e \alpha_m^*] + |\chi|^2)\bar{p} - \text{Im}[\alpha_e \alpha_m^*]\text{Im}[E \times H^*])$$
$$- \frac{2ck^4}{3\pi\omega}(\frac{1}{\varepsilon_0}\text{Re}[\alpha_e \chi^*]\bar{\Phi}_e + \frac{1}{\mu_0}\text{Re}[\alpha_m \chi^*]\bar{\Phi}_m)$$
(8)

$\bar{p} = \frac{1}{2}\text{Re}[E \times H^*]$ is the time average poynting vector; $\bar{\Phi}_E = -\frac{\omega}{4}\varepsilon_0 \text{Im}[E \times E^*]$ and

$\bar{\Phi}_M = -\frac{\omega}{4}\mu_0 \text{Im}[H \times H^*]$ are the electric and magnetic parts of the chirality flow density. The first and second term in (7) is the conservative gradient force and the non-conservative radiation pressure associated with the energy density and energy flow while the third and fourth term are exactly the counterparts in chirality domain. The energy flow $\bar{p}$ and chirality flow $\bar{\Phi}$ are both proportional to the linear momentum of the incident field. In the light-matter interaction, they are transferred to the linear momentum of the particle with different efficiency $\text{Im}[\alpha]$ and $\text{Im}[\chi]$. The last three terms in (7) indicate that the curl of energy/chirality flow could be coupled to linear momentum.

The first term in (8) is proportional to the energy flow. It can be negative for particles within certain ranges, which gives rise to optical pulling force[33]. Besides the electric and magnetic polarizabilities, the chiral property of the object could also modulate this term and the chirality always makes a positive contribution to the optical pulling force since $|\chi|^2$ is always positive. The last term in (8) is the core of this paper and lead to the chiral trapping force. The term indicates that chirality flow of the field could be transferred to the linear momentum of the particle. But being different from $\langle F_{EM} \rangle$, the coupling is done in a way that the electric part $\bar{\Phi}_e$ and magnetic part $\bar{\Phi}_m$ are transferred with different efficiencies which are proportional to $\text{Re}[\alpha_e \chi^*]$ and $\text{Re}[\alpha_m \chi^*]$ respectively.

Substitute the electric field of equation (3) into (7) and (8), we get $\langle F_{EM} \rangle = 0$ since

$\nabla |E|^2 + \nabla |H|^2$, $\bar{p}$, $\bar{K}$ and $\bar{\Phi}$ are all zero. The total force on chiral dipoles can be written as following:

$$\langle F \rangle = \langle F_{INT} \rangle = -\frac{2ck^4}{3\pi\omega}(\frac{1}{\varepsilon_0}\text{Re}[\alpha_e \chi^*]\bar{\Phi}_e + \frac{1}{\mu_0}\text{Re}[\alpha_m \chi^*]\bar{\Phi}_m)$$
$$= \frac{E_{inc}^2 ck_0^4}{3\pi}(-\text{Re}[\alpha_e \chi^*] + \frac{\varepsilon_0}{\mu_0}\text{Re}[\alpha_m \chi^*])\sin(2k_0 x)\hat{x} \quad (9)$$

For an 'ideal' dual symmetric particle with $\frac{\alpha_e}{\varepsilon_0} = \frac{\alpha_m}{\mu_0}$, this force is zero since the electric and magnetic components of the force exactly cancel each other, but in practice they may differ significantly. Within the limit of small particle when the quadrupole and higher order terms are insignificant, we only consider Mie coefficients $a_1$ and $b_1$ for scattering coefficients $\alpha_e$ and $\alpha_m$. We derive $\alpha_e$, $\alpha_m$ and $\chi$ following the method in supplementary material of [23]. Substituting the scattering coefficients into (9), the optical force is calculated on chiral sphere with $\kappa = 0.5$, $\varepsilon_r = 5$ at the position of x = 75nm (where chiral force is maximum in space) with varying the particle's radius from 20nm to 200nm. The electric and magnetic parts of the force are separately plotted is Fig.2a.

The electric and magnetic components of $\bar{\Phi}$ flow with same magnitude and in opposite directions. In the interaction with the chiral particle, they are converted to linear momentum of the particle with different efficiency due to dual-asymmetry of matter and the net momentum leads to the trapping force we observed in pervious section. We also compare the analytically derived force with the full wave simulation result and plotted it in Fig.2b. These results match well in the region of dipole approximations. The trapping effect exists for particles within wide range of permittivity and radius as shown in Fig. 2c and Fig. 2d.

## 5. Three beams configuration

The trapping force on chiral particles can be further extended to more complex 2D cases. The proposed optical field is consisted of one pair of counter-propagating waves with same polarization propagating in y direction and a third wave with orthogonal polarization propagating in x direction (Fig.3a). The intensity of the x direction propagating wave is half of the y direction propagating wave. The waves in y direction have their electric field interference with each other. Any object with electric responses will 'see' the interference pattern and trapped in the dimension. For the third wave, its interferences with the former two waves are conducted in chirality domain instead of energy domain. Only objects with chirality could 'see' the interference pattern and be trapped in 2D space.

We have done the full wave simulations for particles with radius =100nm, $\varepsilon_r =2$ and $\kappa = 0.5, -0.5$ and $0$ when they are placed at different positions in the optical field. Optical forces are calculated by Stress Tensor method [34-37] and two-dimensional force maps are plotted in Fig. 3b, c and d. It can be seen chiral particles with opposite handedness could be trapped at different positions in two-dimensional space while particles without chirality are only trapped in y direction and have very limited force in x direction. It is interesting to find that even for the same optical field, different objects will see (as shown in Fig. 3a, b and c) different potential landscapes. The situation could even extend to 3D cases where the optical field is consisted of 3 orthogonally propagating beams with orthogonal polarizations. In this case, all the interference is conducted in chirality domain and only chiral particles will be trapped in space. These fields could be used to identify, selective trap or filter chiral objects out of normal objects.

## 6. Change in force due to particle size and background Refractive Index

Now, we shall investigate the robustness of the set-ups proposed so far. To make the investigations simpler, we have considered the trapping set-up with two beams instead of the three-beam set-up. However, in this section, we shall apply more rigorous techniques (i.e. Minkowski stress tensor and Lorentz force method) to obtain the time averaged optical force on a generic chiral object even if it is immersed / embedded in a material medium. The 'outside optical force' [38-40] is calculated in this section by the integration of time averaged Minkowski [34-40] stress tensor at $r=a^+$ employing the background fields of the scatterer of radius $a$:

$$\langle \boldsymbol{F}_{\text{Total}}^{\text{Out}} \rangle = \int \langle \bar{\bar{\boldsymbol{T}}}^{\text{out}} \rangle \cdot d\boldsymbol{s},$$
$$\langle \bar{\bar{\boldsymbol{T}}}^{\text{out}} \rangle = \frac{1}{2}\text{Re}[\boldsymbol{D}_{\text{out}}\boldsymbol{E}_{\text{out}}^* + \boldsymbol{B}_{\text{out}}\boldsymbol{H}_{\text{out}}^* - \frac{1}{2}\bar{\bar{\boldsymbol{I}}}(\boldsymbol{E}_{\text{out}}^* \cdot \boldsymbol{D}_{\text{out}} + \boldsymbol{H}_{\text{out}}^* \cdot \boldsymbol{B}_{\text{out}})]$$

(10)

Where 'out' represents the exterior total field of the scatterer; $\boldsymbol{E}$, $\boldsymbol{D}$, $\boldsymbol{H}$ and $\boldsymbol{B}$ are the electric field, displacement vector, magnetic field and induction vectors respectively, $\langle\ \rangle$ represents the time average and $\bar{\bar{I}}$ is the unity tensor. On the other hand, based on the Lorentz force, the total force (surface force and the bulk force [41-44]) can be written as:

$$\langle F_{Total}\rangle = \langle F_{Volume}\rangle = \langle F_{Bulk}\rangle + \langle F_{Surf}\rangle = \int\langle f_{Bulk}\rangle dv + \int\langle f_{Surface}\rangle ds \tag{11}$$

Where

$$\langle f_{Surface}\rangle = [\sigma_e E^*_{avg} + \sigma_m H^*_{avg}]_{at\ r=a}$$
$$= \sigma_e\left(\frac{E_{out}+E_{in}}{2}\right)^*_{at\ r=a} + \sigma_m\left(\frac{H_{out}+H_{in}}{2}\right)^*_{at\ r=a}, \tag{12}$$

$f_{Surface}$ is the surface force density (the force which is felt by the bound electric and magnetic surface charges of a scatterer), which is calculated just at the boundary of a scatterer [41-44]. 'in' represents the interior fields of the scatterer; 'avg' represents the average of the field. $\sigma_e$ and $\sigma_m$ are the bound electric and magnetic surface charge densities of the scatterer respectively, which can be calculated in a similar way shown in [45,46]. The unit vector $\hat{n}$ is the outward pointing normal to the surface. $\varepsilon_b$ is permittivity and $\mu_b$ is permeability of the background.

$$\langle f_{Bulk}\rangle = \frac{1}{2}\text{Re}[\varepsilon_b(\nabla\cdot E_{in})E^*_{in} + \mu_b(\nabla\cdot H_{in})H^*_{in}] - \frac{1}{2}\text{Re}[(i\omega P_{Chiral}\times B^*_{in}) - (i\omega M_{Chiral}\times D^*_{in})] \tag{13}$$

$f_{Bulk}$ is the bulk force density, which is calculated from the interior of the scatterer by employing the inside field [41-44]. In Eq. (13), the effective polarization and magnetization are defined as: $P_{Chiral} = P_e + M_c; P_e = (\varepsilon_s - \varepsilon_b)E_{in}; M_c = i(\kappa/c)H_{in}$ and $M_{Chiral} = M_n + P_c; M_n = (\mu_s - \mu_b)H_{in}; P_c = -i(\kappa/c)E_{in}$ respectively. However, $D^{chiral}_{in}$ and $B^{chiral}_{in}$ in Eq. (13s) should be written directly as: $D^{Chiral}_{in} = \varepsilon_r\varepsilon_0 E_{in} + i\kappa/cH_{in}$ ; $B^{Chiral}_{in} = u_r u_0 H_{in} - i(\kappa/c)E_{in}$. $\varepsilon_s$ is permittivity and $\mu_s$ is permeability of the scatterer. It should be noted that for Mie ranged objects, the dipolar force formula cannot be applied. But the stress tensor method and the Lorentz force formula remain valid for both dipolar and Mie ranged objects [40,41,43,44].

Fig. 5(a) shows the optical force for radius 100nm and 300nm chiral nanoparticle ($\kappa=0.5$). The magnitude of the force for 300nm chiral nanoparticle is much higher compared to 100nm particle but the trapping position almost remains the same. The magnitude of the optical force for chiral nanoparticle mainly originates from scattering of the nanoparticle. Small sphere creates stronger diffraction when light projects on the scatterer. As the radius of the particle increases, the total scattering increases (due to the excitation of electric and magnetic multipoles such as quadrupoles, hexapoles etc in Mie objects [47, 48]) instead of diffraction, which leads to larger force magnitude for larger particle. If we consider the Lorentz force analysis, it is previously reported that: with the increase of particle size, the conducting force increases [49] and hence the total time averaged force on an absorbing object increases. The conducting force arises from the complex polarization and magnetization of an absorbing object. The complex polarization and magnetization exist in a chiral object but in a fully different way as shown in Eq (13). So, this increase of total force for the chiral Mie objects can also be explained based on the Lorentz force distribution similar (but certainly not identical) to ref. [49]. It is important to note that: the time averaged force calculated by stress tensor method in Eq (10) and the Lorentz force in Eq (11) lead to the same total force.

Ultimately, we have demonstrated that the proposed idea is not restricted to dipolar chiral objects only. Chiral Mie objects can also be trapped based on the technique proposed in this article.

The magnitude of the optical force changes with changing the refractive index (RI) of the background medium. The time averaged force has been calculated using the stress tensor method given in Eq (10). Fig. 5(b) shows the change in optical force and trapping position due to change in the RI of the background medium. This is mainly due to the creation of interference field of the scattered light. The interference field is different for different refractive index (RI) of the background medium which leads to change in optical force and trapping position. However, the number of trapping positions remain the same (as the air background) when the background medium is water. As most of the experimental configurations usually use the water medium as background, we may conclude that: the proposals presented in this article may remain valid in realistic experimental set-ups with water or other similar background environments.

## 7. Changing of Trapping position due to the change in frequency

The trapping force comes from the difference in chirality-linear momentum transferring efficiency between the electric part and magnetic part due to dual electric-magnetic asymmetry of light-matter interaction. Notably, by considering equilibrium positions, particles with opposite handedness are trapped at different locations separated by a quarter wavelength.

The total force on chiral dipole can be found in Eq (6). Where $k_0$ = wave number and can be written as $k_0 = 2\pi \frac{1}{\lambda} = 2\pi \frac{v}{c}$. Hence, the trapping force intensity and the trapping position varies with the frequency ($v$) which happens for the lambda of electromagnetic radiation. From Fig. 6 we can observe that increase in lambda (decrease of the source energy) causes decrease of wave number. Ultimately, number of the trapping position decreases with increasing the wavelength. The blue [800 nm] and pink [400nm] lines clearly show the change in trapping position with changing the wavelength. It should be noted that the time averaged force has been verified using the rigorous stress tensor method given in Eq (10).

So, we can say that by changing the wavelength we can get tunable trapping position for chiral nanoparticle. Notably, by increasing the energy of incident light (i.e. source frequency), number of the trapping positions can be increased.

## 8. Conclusions

In conclusion, we have theoretically investigated the optical force on chiral particles in optical field consisted of two counter-propagating orthogonally polarized waves. We found a trapping force on chiral particles but not on normal particles. Based on this force, chiral particles could be trapped and differentiated from particles without chirality in one dimensional, two dimensional and three-dimensional space. The trapping force comes from the difference in chirality-linear momentum transferring efficiency between the electric part and magnetic part due to dual electric-magnetic asymmetry of light-matter interaction. We have also shown the tunability in chiral trapping position by varying the wavelength. Notably, by increasing the energy of incident light (i.e. source frequency), number of the trapping positions can be increased. We have demonstrated that the proposed idea is not restricted to dipolar chiral objects only. Chiral Mie objects can also be trapped based on the technique proposed in this article. However, the number of trapping positions remain the same (as the air background) when the background medium is water. As most of the experimental configurations usually use the water medium as background, we may conclude that: the proposals presented in this article may remain valid in realistic experimental set-ups with water or other similar background environments. These results provide new insights into chiral light-matter interactions and may bring out new methods for manipulating chiral objects.

**Figures and captions**

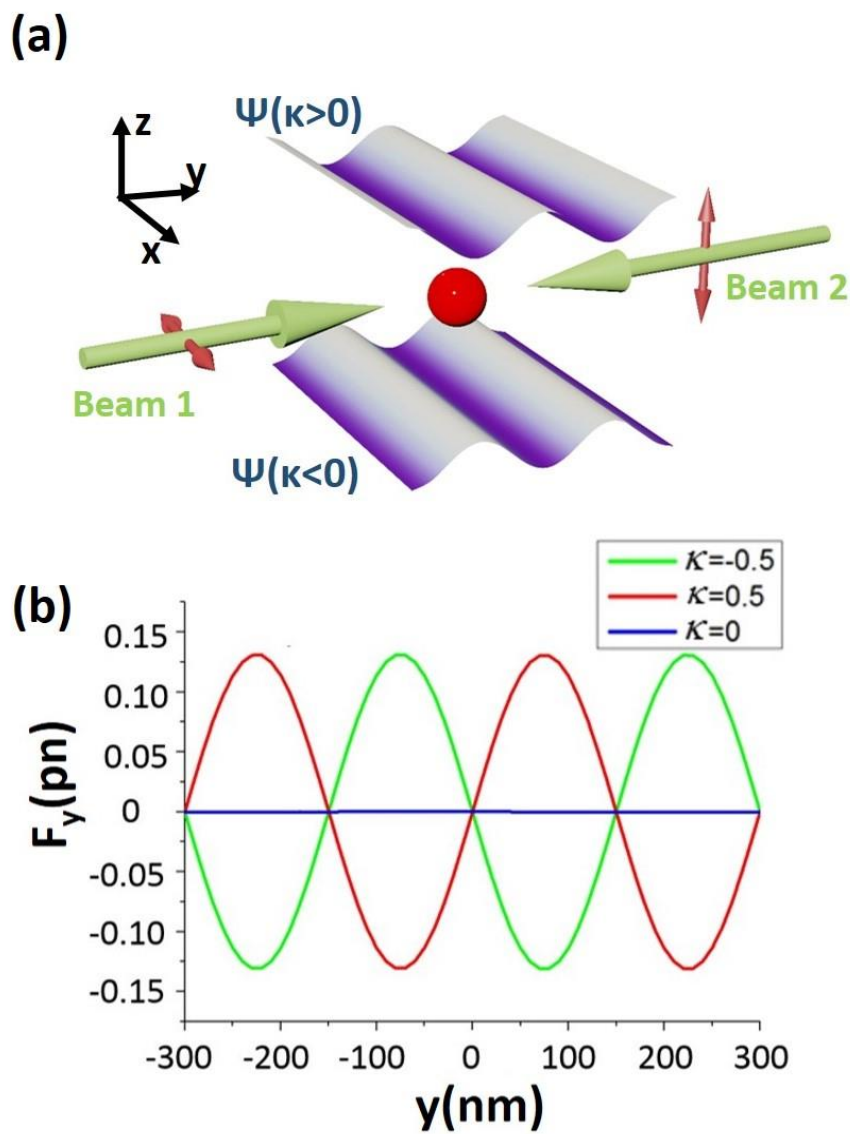

Fig.1 (a) Schematic of the optical field described in current work. It is consisted of two counter-propagating orthogonally polarized waves. Surfaces stand for optical potential for right-handed and left handed chiral particles (b) Numerical calculated optical force on left-handed, right-handed chiral particles and normal particles placed in optical field.

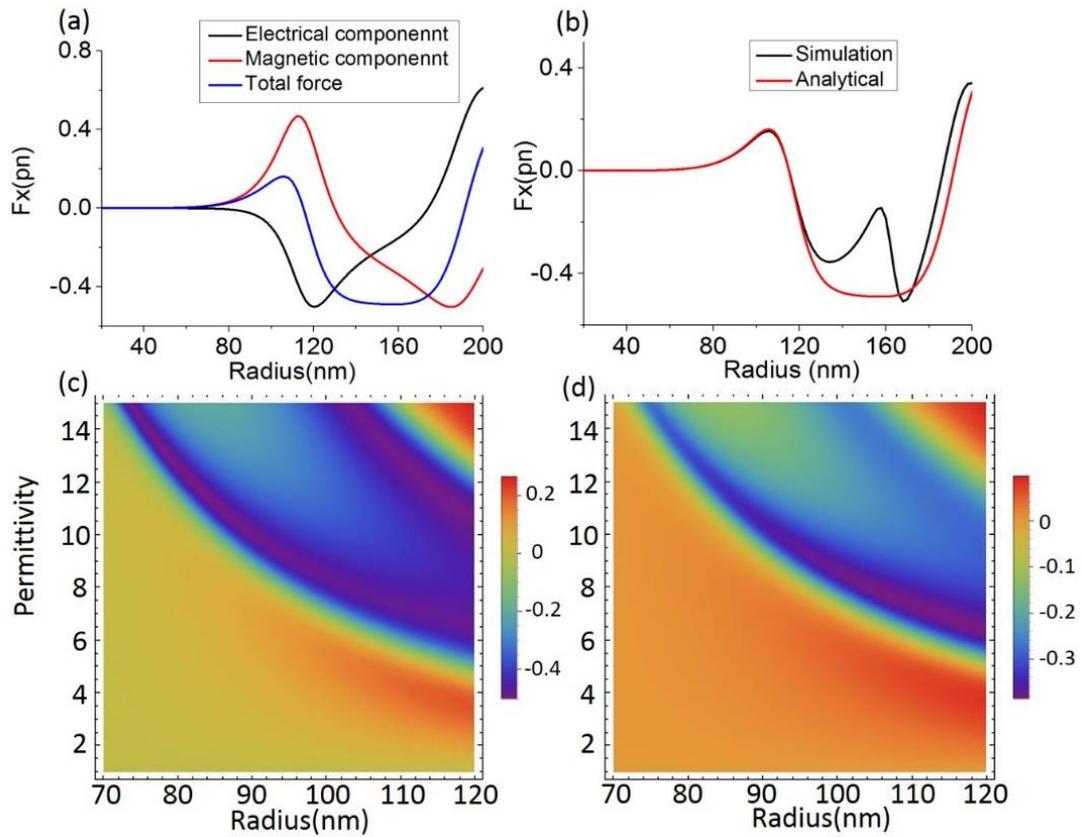

Fig.2 (a) Optical force on a chiral particle derived from analytical theory, the net force is sum of the forces from electric component and magnetic component. (b) The calculated result by analytical theory matches well with the numerical simulated results when the size of the particle is within dipole region. (c) Optical force on chiral particles with different permittivity and radius with chiral parameter $\kappa$=0.5, (d) $\kappa$=0.2.

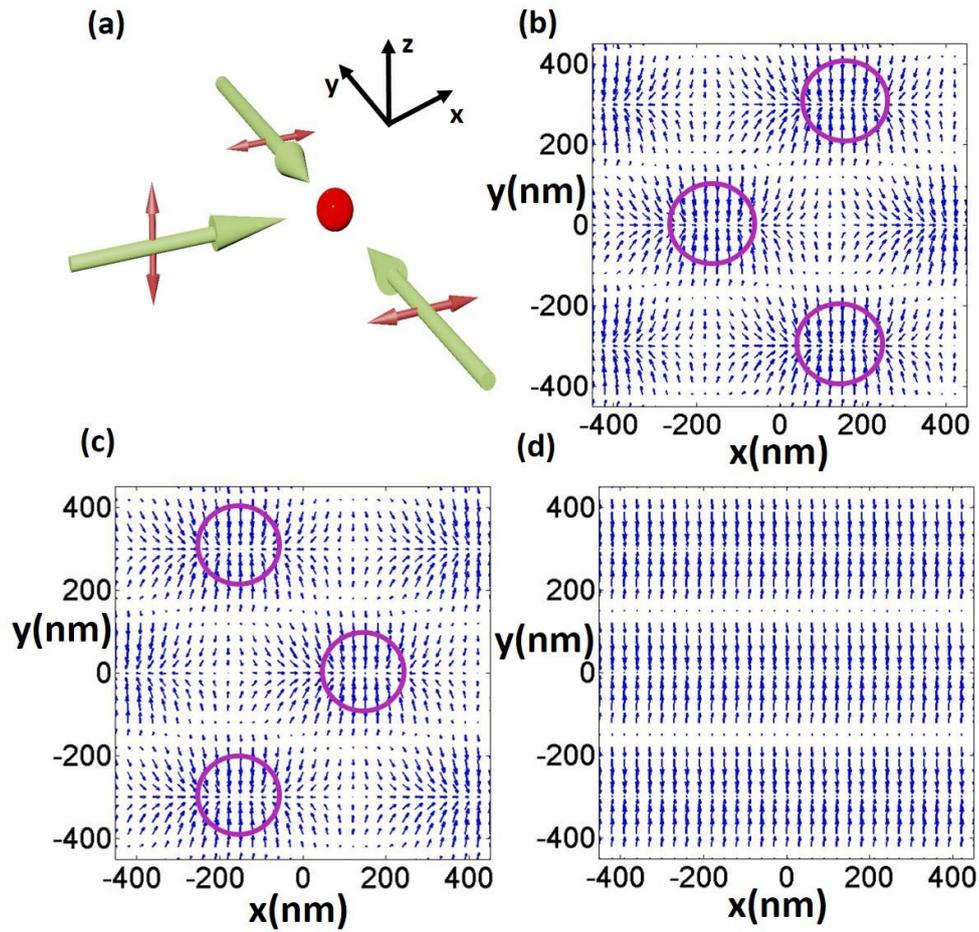

Fig.3 (a) Optical field formed by one standing wave and a third wave with orthogonal polarization inserted perpendicularly. (b) Force map on particles in two-dimensional space for $\kappa=0.5$ (c) $\kappa=-0.5$ and (d) $\kappa=0$. The particles are trapped within the red circle.

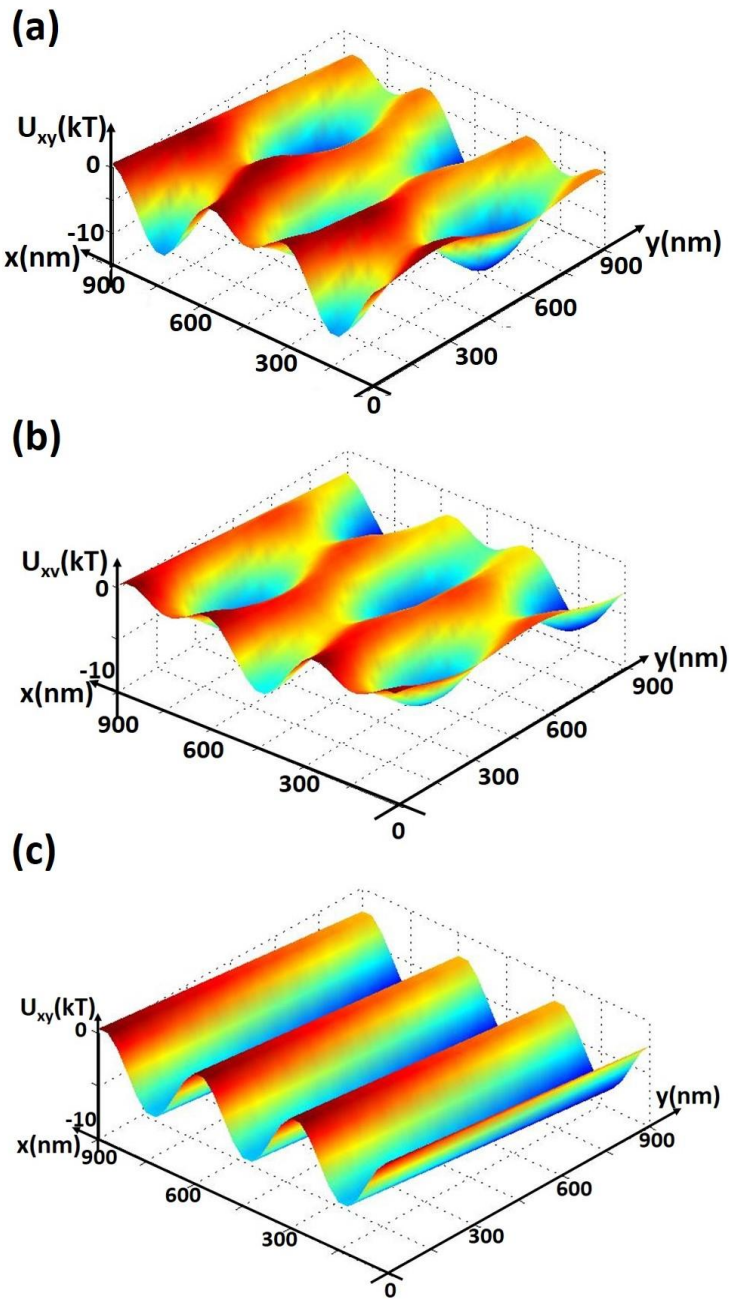

Fig .4 Optical field potential formed by one standing wave and a third wave with orthogonal polarization inserted perpendicularly: Force potential map on particles for (a) $\kappa=0.5$ (b) $\kappa=-0.5$ and (c) $\kappa=0$

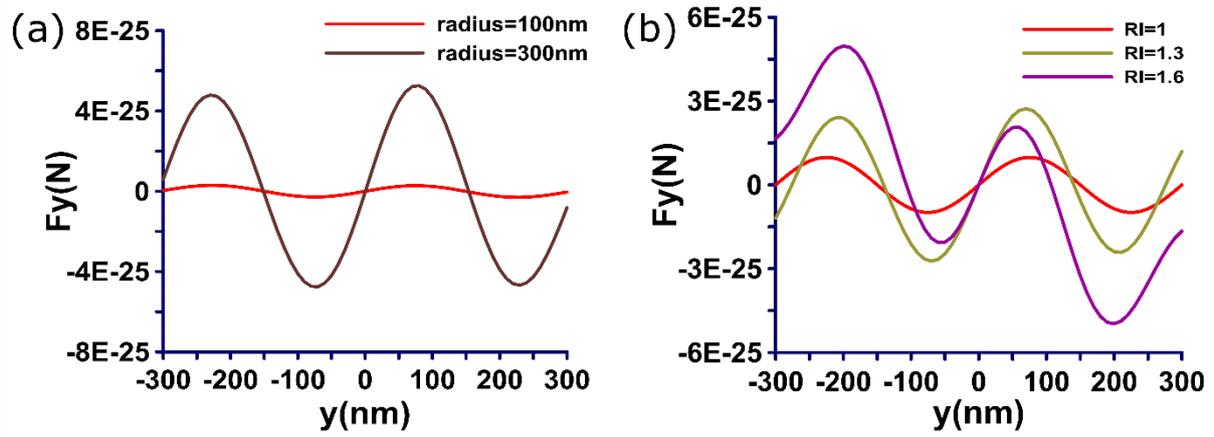

Fig .5 (a) Optical Force for Chiral particle of radius 100nm and 300nm with $\kappa$ =0.5 at wavelength 600 nm in air. (b) Change in Optical force on chiral particle for radius 100nm and $\kappa$ =0.5 due to the change in Refractive Index (n=1, n=1.3) of the medium.

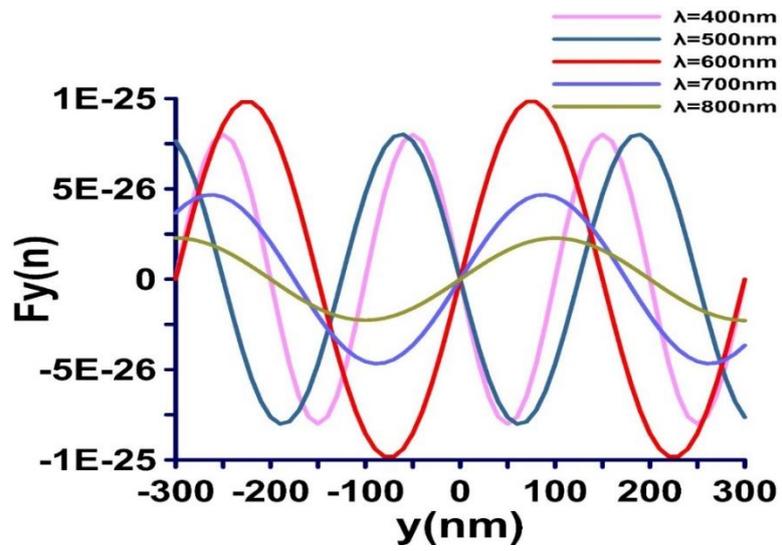

Fig .6 Optical force and Trapping position due to the change in wavelength ( $\lambda = 400nm, 500nm, 600nm, 700nm, 800nm$) of incident wave on chiral particle for $\kappa$ =0.5 and radius 100nm.